\RequirePackage{fix-cm}
\documentclass[smallextended]{svjour3}       
\smartqed  
\usepackage{graphicx,color}
\usepackage{bm}       
\usepackage{bbm}
\usepackage{amsmath}
\usepackage{siunitx}
\begin{document}

\title{Nonlocal structure of the leading order \textit{ab initio} effective potentials for proton elastic scattering
from light nuclei 
}

\titlerunning{pA Effective Potentials}        

\author{Ch.~Elster         \and
        M.~Burrows \and
        R.B.~Baker \and
        S.P.~Weppner \and
        K.D.~Launey \and
        P.~Maris  \and
        G.~Popa
}

\authorrunning{Ch. Elster et {\it al} } 

\institute{Ch. Elster, R.B. Baker, G. Popa \at
  Institute of Nuclear and Particle Physics,  and
   Department of Physics and Astronomy,  Ohio University, Athens, OH 45701, USA  \\
              \email{elster@ohio.edu}     
           \and
           M. Burrows, K.D. Launey \at
             Department of Physics and Astronomy, Louisiana State University,
       Baton Rouge, LA 70803, USA\\
       \and
    S.P. Weppner \at
   Natural Sciences, Eckerd College, St. Petersburg, FL 33711, USA \\
   \and
   P. Maris \at
  Department of Physics and Astronomy, Iowa State University, Ames, IA 50011, USA \\
}

\date{Received: date / Accepted: date}

\maketitle

\begin{abstract}
Based on the spectator expansion of the multiple scattering series we employ a chiral next-to-next-to-leading
order (NNLO) nucleon-nucleon interaction on the same footing in the structure as well as in the reaction
calculation to obtain an in leading-order consistent effective potential for nucleon-nucleus elastic
scattering, which includes the spin of the struck target nucleon. As an example we present proton scattering
off $^{12}$C. 

\end{abstract}

\section{Introduction}
\label{intro}
Recent developments of the nucleon-nucleon (NN) and three-nucleon (3N) interactions, derived
from chiral effective field theory, have yielded major
progress~\cite{EntemM03,Epelbaum06,Epelbaum:2008ga}. These, together with the
utilization of massively parallel computing resources (e.g.,
see~\cite{LangrSTDD12,CPE:CPE3129,Jung:2013:EFO}), have placed {\it ab initio}
large-scale simulations at the frontier of nuclear structure and reaction explorations. 
Here we focus on the use of the \textit{ab initio} no-core shell model (NCSM)
\cite{Navratil:2000ww,Roth:2007sv,BarrettNV13} and symmetry-adapted no-core shell model (SA-NCSM)
\cite{LauneyDD16,Dytrych:2020vkl} to provide the relevant structure inputs. 

Our approach
to elastic nucleon-nucleus (NA) scattering is based on the spectator expansion of multiple scattering
theory. Specifically,  the leading order term in this expansion involves two-body interactions between the projectile and one of the target nucleons which requires a
convolution of fully off-shell NN scattering amplitudes with the nuclear wave functions of
the target represented by a nonlocal one-body nuclear density.  
\begin{figure*}
  \includegraphics[width=0.50\textwidth]{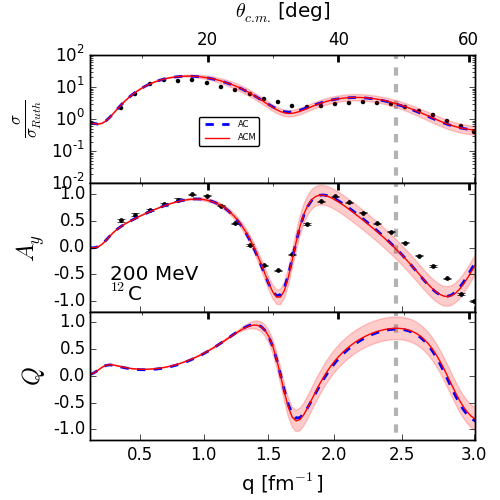}
  \includegraphics[width=0.50\textwidth]{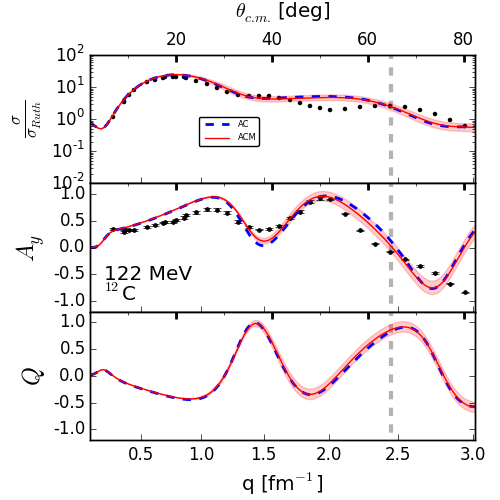}
\caption{Elastic scattering observables for proton scattering off $^{12}$C at 200~MeV (left panel) and 122~MeV
(right panel) projectile energy. Shown are the differential cross section divided by the Rutherford cross
section, the analyzing power $A_y$ and the spin-rotation function $Q$ as function of the c.m. scattering angle
$\theta_{c.m.}$as well as of the momentum transfer $q$. The grey dashed line represents the momentum transfer
up to which the NNLO${_{\rm{opt}}}$ chiral interaction~\protect\cite{Ekstrom13} entering the calculations
is fit to NN scattering data. The red solid line (ACM) represents the consistent leading order {\it ab
initio} calculation of the effective interaction, while for the blue dashed line (AC) the spin of the struck
target nucleon is ignored. The NCSM calculations use $N_{\rm max}$=10 and
$\hbar\omega$=20. The red band represents the variation in the scattering observables when
$\hbar\omega$=16 and 24 MeV are employed. 
The data are taken from Refs.~\cite{Meyer:1981na} for 200~MeV and \cite{Meyer:1983kd}
for 122~MeV. This figure is taken from Ref.~\cite{BurrowsM:2020}.
}
\label{fig:0}       
\end{figure*}

\section{Nucleon-Nucleus effective interaction}
\label{sec:1}
The standard approach to elastic scattering of a strongly interacting projectile from a target of $A$
particles using a multiple scattering approach is a separation of the Lippmann-Schwinger equation for the
transition amplitude $T=V+V G_0(E) T$ into two parts, namely an integral equation for $T$ and one for the
effective potential $U$ in the problem
\begin{equation}
T = U + U G_0(E) P T ~~~~{\rm and}~~~~ U = V+VG_0 QT .
\label{eq1}
\end{equation}
In the above equations the operator $V=\sum_{i=1}^A v_{0i}$ consists of
the NN potential $v_{0i}$ acting between the projectile and
the $i$th target nucleon. The free propagator $G_0(E)$ for the
projectile$+$target system is given by $G_0(E)=(E-H_0+i\epsilon)^{-1}$, and the Hamiltonian for the (A+1)
particle system by $H=H_0+V$. Here the free Hamiltonian is given by
$H_0=h_0+H_A$, where $h_0$ is the kinetic energy operator for the
projectile and $H_A$ stands for the target Hamiltonian. Defining
$|\Phi_{A}\rangle$ as the ground state of the target, we have
$H_A |\Phi_{A}\rangle = E_A |\Phi_{A}\rangle$. The operators $P$ and $Q$ are projection operators, $P+Q=1$. 
When considering elastic scattering $P$ is defined such that $[G_0,P]=0$. 

\begin{figure*}[t]
  \includegraphics[width=0.50\textwidth]{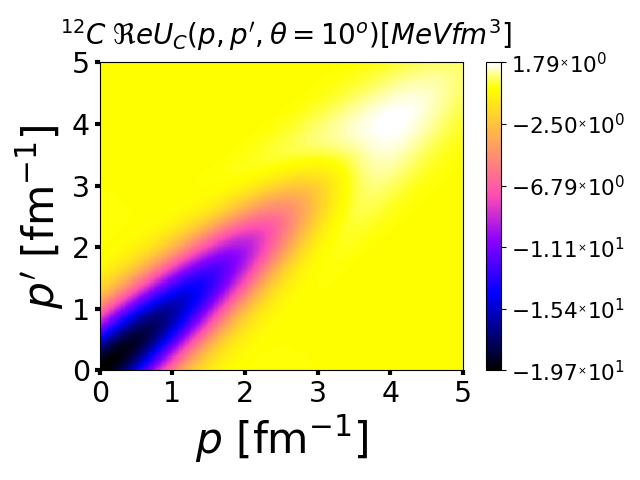}
  \includegraphics[width=0.49\textwidth]{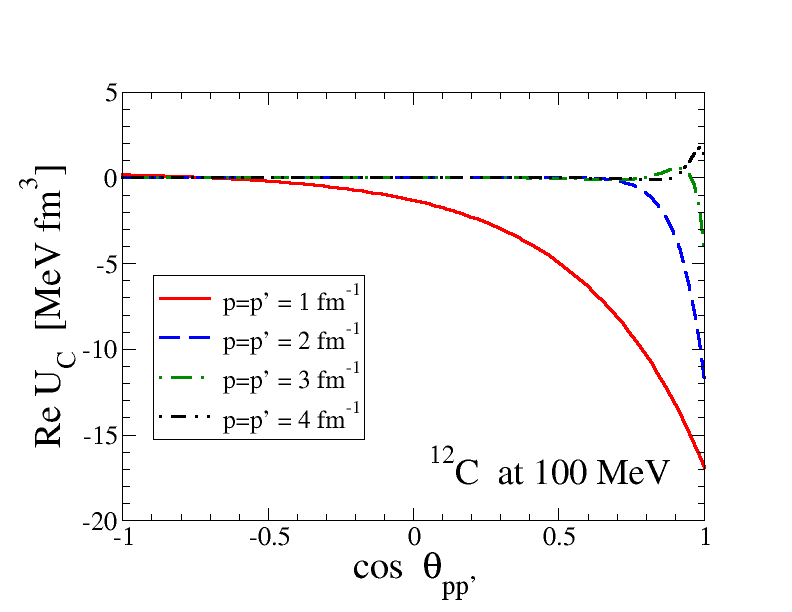}
\caption{The left panel shows the central effective potential for proton scattering off $^{12}$C 
calculated for 100 MeV
projectile energy as function of the momenta $p$ and $p'$ for fixed angle $\theta = 10^o$. 
The potential is based on the NNLO${_{\rm{opt}}}$ chiral
interaction~\protect\cite{Ekstrom13}. The structure calculation employs   
$N_{\rm max}$=10 and $\hbar\omega$=20. The right panel depicts the angle dependence of the central effective
potential for fixed values of $p=p'$ given in the legend.}
\label{fig:1}
\end{figure*}

The fundamental idea of the spectator expansion for the effective interaction is an ordering of the scattering
process according to the number of active target nucleons interacting directly with the projectile. Thus in
leading order only two active nucleons are considered.  For the leading order term being {\it ab initio} means
that the NN interaction for the active pair is considered at the same level as the NN interaction employed to
obtain the ground state wave functions $|\Phi_{A}\rangle$. 
Detailed aspects of the derivation of the leading order term and how the spin structure of the NN interaction
is consistently taken into account in the reaction process are given in
Refs.~\cite{Burrows:2018ggt,Burrows:2020qvu,BurrowsM:2020}. The resulting final 
expression for the {\it ab initio} leading order effective interaction is given as~\cite{Baker:2021izp}
%
\begin{eqnarray}
\label{eq2}
\lefteqn{\widehat{U}_{\mathrm{p}}(\bm{q},\bm{\mathcal{K}}_{NA},\epsilon) =} &&   \\
& & \sum_{\alpha=\mathrm{n,p}} \int d^3{\mathcal{K}} \eta\left( \bm{q}, \bm{\mathcal{K}},
\bm{\mathcal{K}}_{NA} \right)
A_{\mathrm{p}\alpha}\left( \bm{q}, \overline{\bm{\mathcal{K}_A}} ; \epsilon
\right) \rho_\alpha^{K_s=0} \left(\bm{\mathcal{P}'}, \bm{\mathcal{P}}  \right) \cr
&+&i (\bm{\sigma^{(0)}}\cdot\hat{\bm{n}}) \sum_{\alpha=\mathrm{n,p}} \int d^3{\mathcal{K}} \eta\left( \bm{q},
\bm{\mathcal{K}}, \bm{\mathcal{K}}_{NA} \right)
C_{\mathrm{p}\alpha}\left( \bm{q}, \overline{\bm{\mathcal{K}_A}} ; \epsilon
\right) \rho_\alpha^{K_s=0} \left(\bm{\mathcal{P}'}, \bm{\mathcal{P}}  \right) \cr
&+&i \sum_{\alpha=\mathrm{n,p}} \int d^3{\mathcal{K}} \eta\left( \bm{q}, \bm{\mathcal{K}},
\bm{\mathcal{K}}_{NA}
\right) C_{\mathrm{p}\alpha} \left( \bm{q}, \overline{\bm{\mathcal{K}}_A}; \epsilon \right)
 S_{n,\alpha} \left(\bm{\mathcal{P}'}, \bm{\mathcal{P}} \right) \cos \beta\cr
&+&i (\bm{\sigma^{(0)}}\cdot\hat{\bm{n}}) \sum_{\alpha=\mathrm{n,p}} \int d^3{\mathcal{K}} \eta\left( \bm{q},
\bm{\mathcal{K}}, \bm{\mathcal{K}}_{NA} \right)  (-i)
M_{\mathrm{p}\alpha} \left( \bm{q}, \overline{\bm{\mathcal{K}}_A} ; \epsilon \right)
 S_{n,\alpha} \left(\bm{\mathcal{P}'}, \bm{\mathcal{P}}  \right) \cos \beta  . \nonumber
\end{eqnarray}
%
The term $\eta\left( \bm{q}, \bm{\mathcal{K}}, \bm{\mathcal{K}_{NA}} \right)$  is the M{\o}ller
factor~\cite{CMoller} describing the transformation from the NN frame to the NA frame.
The functions $A_{\mathrm{p}\alpha}$, $C_{\mathrm{p}\alpha}$, and $M_{\mathrm{p}\alpha}$ represent the NN
amplitudes in the Wolfenstein
representation~\cite{wolfenstein-ashkin}. Since the incoming proton can interact
with either a proton or a neutron in the nucleus, the index $\alpha$ indicates the
neutron ($\mathrm{n}$) and proton ($\mathrm{p}$) contributions, which are calculated separately and then
summed up.
With respect to the nucleus, the operator $i (\bm{\sigma^{(0)}}\cdot \hat{\bm{n}})$ represents the momentum
space spin-orbit
operator of the projectile. As such, Eq.~(\ref{eq2}) exhibits the
expected form of an interaction between a spin-$\frac{1}{2}$ projectile and a target nucleus in a $J=0$ state
\cite{RodbergThaler}.
The momentum vectors in the problem are given as
\begin{eqnarray}
\label{eq3}
\bm{q} &=& \bm{p'} - \bm{p} = \bm{k'} - \bm{k}, ~~~ 
\bm{\mathcal{K}} = \frac{1}{2} \left(\bm{k'} + \bm{k}\right),  ~~~
\hat{\bm{n}}=\frac{\bm{\mathcal{K}} \times \bm{q}}{\left| \bm{\mathcal{K}} \times
\bm{q}\right|} \cr
\bm{\mathcal{K}_{NA}} &=& \frac{A}{A+1}\left[\left(\bm{p'} + \bm{p}\right) +
      \frac{1}{2} \left(\bm{k'} + \bm{k}\right) \right], ~~~
\overline{\bm{\mathcal{K}}_A}  =  \frac{1}{2}\left( \frac{A+1}{A}\bm{\mathcal{K}}_{NA} -
\bm{\mathcal{K}}\right), \cr
\bm{\mathcal{P}}&=& \bm{\mathcal{K}}+\frac{A-1}{A}\frac{\bm{q}}{2}, ~~~
\bm{\mathcal{P'}} =  \bm{\mathcal{K}}-\frac{A-1}{A}\frac{\bm{q}}{2}  .
\end{eqnarray}
The momentum dependence in Eq.~(\ref{eq2}) is quite intricate: Although the potential is needed in the
(A+1)-body frame, its ingredients are derived in the NN frame as well as in the frame of the nucleus.
A derivation and
calculation based on the momentum transfer ${\bm{q}}$ and the momentum 
$\bm{\mathcal{K}_{NA}}$ are convenient
due to the invariance of ${\bm q}$ in all frames. However, these variables may not be as intuitive for
visualizing the potential. In this short contribution we want to concentrate on elastic proton scattering from
$^{12}$C using an {\it ab initio} effective potential based on the  
 NNLO${_{\rm{opt}}}$ chiral interaction~\cite{Ekstrom13} and a NCSM structure calculation based on the same
interaction. 
%
\begin{figure*}[b]
  \includegraphics[width=0.50\textwidth]{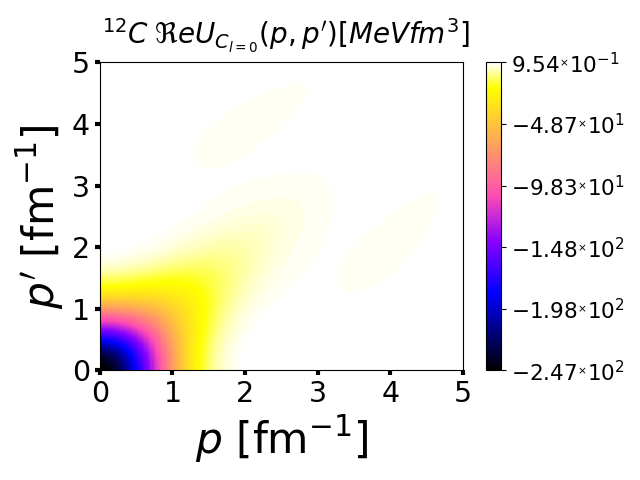}
  \includegraphics[width=0.50\textwidth]{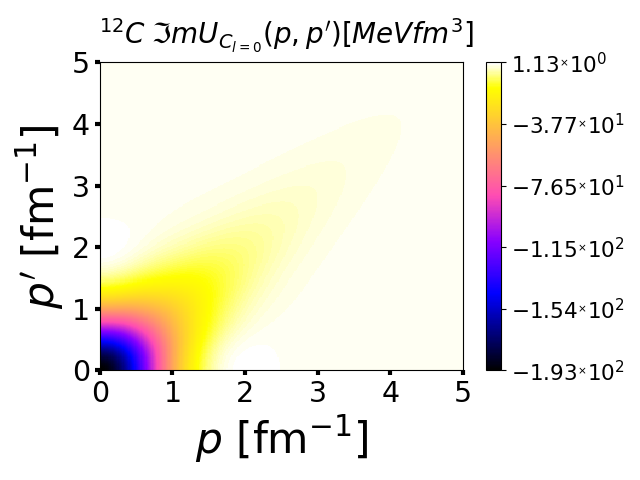}
\caption{The real (left panel) and imaginary (right panel)  s-wave central effective potential 
for proton scattering off $^{12}$C calculated for 100 MeV
projectile energy as function of the momenta $p$ and $p'$. The calculation is based on the same input as the
one of Fig.~\ref{fig:1}.}
\label{fig:2}       
\end{figure*}
\begin{figure*}
  \includegraphics[width=0.50\textwidth]{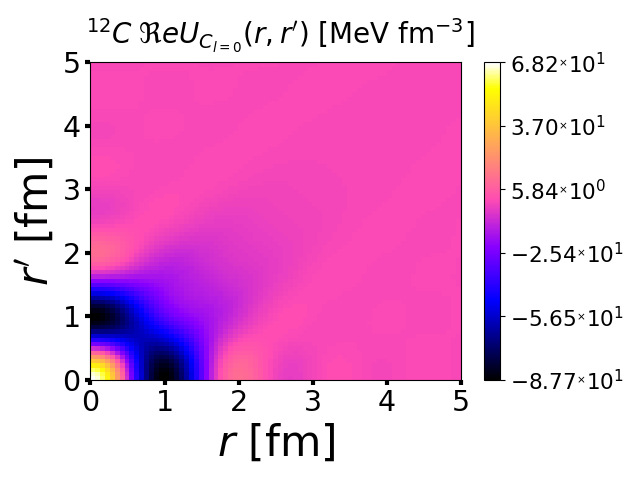}
  \includegraphics[width=0.50\textwidth]{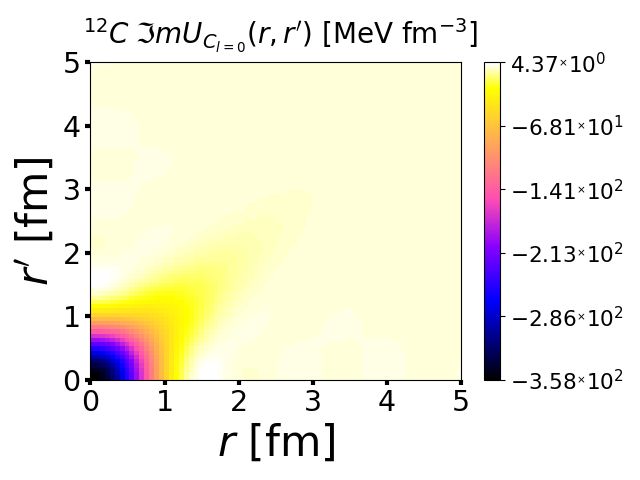}
\caption{Same as Fig.~\ref{fig:2} but for the coordinate space s-wave central effective potential.
}
\label{fig:3}       
\end{figure*}

In Fig.~\ref{fig:0} the elastic scattering observables, $d\sigma/d\Omega$ divided by the Rutherford cross
section, analyzing power and spin rotation function, are shown for two different projectile energies and
compared to existing data.  The differential cross sections are very well described up to $q \sim$2~fm$^{-1}$, as is
the analyzing power at 200~MeV, while at 122~MeV it is slightly over-predicted at smaller $q$. 

The effective interaction from Eq.~(\ref{eq2}) is nonlocal as well as energy dependent, and depends on two
vector momenta. In order to visualize
it in a more traditional form used in solving a momentum-space Lippmann-Schwinger equation, Fig.~\ref{fig:1}
depicts the real central part of the effective interaction as function of the momenta $p$ and $p'$ and an
angle $\theta_{pp'} =10^o$ together with the angle dependence at four different $p=p'$ values. The
effective interaction is clearly peaked in forward direction. 
For Fig.~\ref{fig:2} we perform a partial wave decomposition of the central part and show the s-wave
contribution as function of $p$ and $p'$. A Fourier transform leads to coordinate space representation of
the central s-wave potential shown in Fig.~\ref{fig:3}. The radius obtained from the corresponding NCSM calculation
of the ground state is about 2~fm. This is consistent with the figure showing the potential has its largest
contributions for $r=r' \le$2~fm. It is also worthwhile to note that the largest negative contributions to
the potential are not located along the line $(r+r')/2$. 

\section{Summary}

In this short contribution we concentrate on elastic proton scattering off $^{12}$C and a visualization of the
{\it ab initio} effective interaction entering the calculation. The effective interaction is complex,
nonlocal and a function of two vector momenta and the energy. We show the s-wave projected central part of the
interaction in momentum as well as in coordinate space. 
The nonlocal structure of the real central s-wave effective potential for $^{12}$C is quite similar in
structure to the one described in Ref.~\cite{Arellano:2018jjd} for the heavier nucleus $^{40}$Ca, which is
based on folding with a two-nucleon g-matrix.  We also
want to point out that the overall structure of the real s-wave potential from Fig.~\ref{fig:3} has very
similar characteristics to a real s-wave potential calculated from a Green's function approach combined with
the coupled cluster approach for the nucleus $^{16}$O~\cite{Rotureau:2016jpf}.

\begin{acknowledgements}
This work was performed in part under the auspices of the U.~S. Department of Energy under contract Nos.
DE-FG02-93ER40756 and DE-SC0018223, and by the U.S. NSF (OIA-1738287 \& PHY-1913728). The numerical
computations benefited from computing resources provided by Blue Waters (supported by the U.S. NSF,
OCI-0725070 and ACI-1238993, and the state of Illinois),  the Louisiana Optical Network Initiative
and HPC resources provided by LSU, and resources of the National Energy Research Scientific
Computing Center, a DOE Office of Science User Facility supported by the Office of Science of the U.S.
Department of Energy under contract No. DE-AC02-05CH11231.
\end{acknowledgements}

\bibliographystyle{spphys}       
\bibliography{clusterpot,denspot,ncsm}   

\end{document}